\documentclass[twocolumn, trackchanges]{aastex63}
\newcommand{\lasp}{
	Laboratory for Atmospheric and Space Physics\\
	University of Colorado at Boulder\\
	3665 Discovery Dr.\\
	Boulder, CO 80303-7814, USA
	}
\newcommand{\cires}{ 
	Cooperative Institute for Research in Environmental Sciences\\
	University of Colorado at Boulder\\
	Boulder, CO 80309-0390, USA
	}
\newcommand{\noaa}{ 
	NOAA National Centers for Environmental Information\\
	Boulder, CO 80305, USA
	}

\newcommand{\nShort}{10}
\newcommand{\nLong}{4}
\newcommand{\tShort}{0.035}
\newcommand{\tLong}{15}

\newcommand{\Rs}{\nom{R}}
\usepackage{textcomp}
\usepackage{gensymb}

\received{2021 July 6}
\revised{2021 September 10}
\accepted{2021 $<$insert$>$}

\submitjournal{AJ} % Submitting to LABORATORY ASTROPHYSICS, INSTRUMENTATION, SOFTWARE, AND DATA corridor

\shorttitle{Simultaneous High Dynamic Range}
\shortauthors{Mason et al.}

\graphicspath{{./}{}}

\begin{document}

\title{Simultaneous High Dynamic Range Algorithm, Testing, and Instrument Simulation}

\correspondingauthor{James Paul Mason}
\email{james.mason@lasp.colorado.edu}

\author[0000-0002-3783-5509]{James Paul Mason}
\affiliation{\lasp}

\author[0000-0002-0494-2025]{Daniel B. Seaton}
\affiliation{\cires}
\affiliation{\noaa}

\author[0000-0001-5533-5498]{Andrew R. Jones}
\affiliation{\lasp}

\author[0000-0002-9672-3873]{Meng Jin}
\affiliation{Lockheed Martin Solar \& Astrophysics Laboratory}
\affiliation{SETI Institute}

\author[0000-0003-4372-7405]{Phillip C. Chamberlin}
\affiliation{\lasp}

\author[0000-0002-4546-2394]{Alan Sims}
\affiliation{\lasp}

\author[0000-0002-2308-6797]{Thomas N. Woods}
\affiliation{\lasp}

\begin{abstract}

Within an imaging instrument's field of view, there may be many observational targets of interest. Similarly, within a spectrograph's bandpass, there may be many emission lines of interest. The brightness of these targets and lines can be orders of magnitude different, which poses a challenge to instrument and mission design. A single exposure can saturate the bright emission and/or have a low signal to noise ratio (SNR) for faint emission. Traditional high dynamic range (HDR) techniques solve this problem by either combining multiple sequential exposures of varied duration or splitting the light to different sensors. These methods, however, can result in the loss of science capability, reduced observational efficiency, or increased complexity and cost. The simultaneous HDR method described in this paper avoids these issues by utilizing a special type of detector whose rows can be read independently to define zones that are then composited, resulting in areas with short or long exposure measured simultaneously. We demonstrate this technique for the sun, which is bright on disk and faint off disk. We emulated these conditions in the lab to validate the method. We built an instrument simulator to demonstrate the method for a realistic solar imager and input. We then calculated SNRs, finding a value of 45 for a faint coronal mass ejection (CME) and 200 for a bright CME, both at 3.5 \Rs -- meeting or far exceeding the international standard for digital photography that defines a SNR of 10 as acceptable and 40 as excellent. Future missions should consider this type of hardware and technique in their trade studies for instrument design.

\end{abstract}


\section{Introduction} 
\label{sec:intro}

Many observational targets present high dynamic range in brightness. Spectral emission lines within a particular wavelength range of interest are not obligated to be similarly intense. For example, the Hubble Space Telescope observes Lyman $\alpha$\ (1216 \AA) which is many times brighter than nearby spectral lines it also measures such as \ion{Si}{3} (1206 \AA), \ion{O}{1} (1304 \AA), and \ion{C}{2} (1335 \AA). These lines are used for spectroscopy of transiting exoplanets and their host stars but the dynamic range requires that the dimmer lines be observed separately, necessitating multiple Hubble orbits (e.g., \citealt{Munoz2020, Munoz2021}) -- this reduces the observation efficiency of a highly oversubscribed telescope as well as complicates the scientific analysis, which would prefer simultaneous measurements of all of these emission lines. 

Similarly, within a particular instrument field of view, various objects may have wildly different intensities. The disk of the sun is much brighter than the off-disk corona: 10$^9$ times brighter in white light by 2.5 nominal solar radii (\Rs) and 10$^4$ times brighter in the extreme ultraviolet (EUV; \citealt{Golub2010, Seaton2021}). This is why observation of the middle ($\geq \sim$1.5 \Rs) and high ($\geq \sim$5 \Rs) corona is typically achieved by occulting the solar disk, for example during solar eclipses or with coronagraphs. An unocculted exposure long enough to capture the corona leads to a completely saturated disk that could result in detector blooming that masks part of the corona. Moreover, scattered light is a problem that must be solved for any solar imager interested in the off-disk corona: even a small percentage of the copious disk photons scattering into the off-disk part of the image drives the signal to noise ratio (SNR) very low. Any science questions targeting a phenomenon that cross from on disk to off (e.g., coronal mass ejections), therefore, requires the combination of multiple optimized instruments. This introduces challenges with cross-calibration and coordination, as well as caveats in the conclusions drawn if, for example, different physical structures are being compared due to the observations being in different wavelength regimes. Such is the case with coronal mass ejections (CMEs) observed on disk with EUV but with white light off disk.

\begin{figure*}
    \plotone{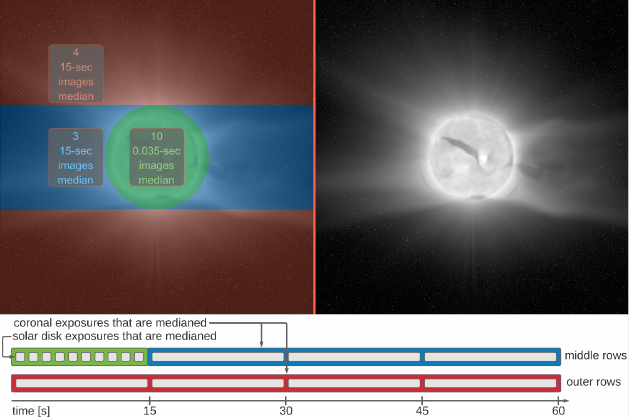}
    \caption{(left) Annotated depiction of the readout scheme, with detector regions highlighted. (right) Synthetic SunCET image composite without colored highlights. Both images are produced by the SunCET instrument simulator (Section \ref{sec:instrument_simulator}). (bottom) Flowchart of algorithm.
    \label{fig:shdr_median_algorithm}}
\end{figure*}

High dynamic range (HDR) imaging solves these problems but can introduce new ones. In traditional HDR, varying exposures are combined from either sequential images or separate sensors \citep{Unger2016}. Problems remain with these methods if 1) important temporal dynamics are lost or 2) a multi-sensor solution is prohibitively complex or expensive. Time dependence is a critical component of both examples in the previous paragraphs: exoplanet transits are transient phenomena that contain important atmospheric information in their time profiles and coronal mass ejections constrain plasma particle acceleration models through their time-dependent kinematic profiles (\citealt{Mason2021}, and references therein). Multiple sensors necessitate either beam splitters or multiple telescopes. Splitters reduce the number of photons and therefore the SNR, which is likely to majorly constrain HDR-oriented instruments because they are intended to measure faint sources. The alternative of employing multiple telescopes drastically drives up the cost of the observatory. Both options also increase the complexity of the design, necessitating additional mechanical, electrical, and/or software interfaces that, in turn, requires additional testing. 

We present a method to achieve \emph{simultaneous} HDR imaging with a single sensor through spatial partitioning where each region integrates for an independent duration. Section \ref{sec:shdr} describes the algorithm, Section \ref{sec:lab} presents our laboratory validation, and Section \ref{sec:instrument_simulator} details a software instrument simulator for the coronal mass ejection science case described above.

\section{Simultaneous High Dynamic Range Algorithm}
\label{sec:shdr}

Our Simultaneous High Dynamic Range (SHDR) algorithm only requires a detector whose rows can be read out independently. We have identified several Complementary Metal Oxide Semiconductor (CMOS) detectors with this capability and have employed it with one: a Teledyne e2v CIS115. The CIS115 detector is designed such that all the pixels in a single row are sampled at the same time. Rows are individually selected through a parallel address bus that is controlled by Field-Programmable Gate Array (FPGA) logic in our custom readout electronics. Only the selected row has its pixels sampled and reset. We developed the SHDR algorithm to support the Sun Coronal Ejection Tracker (SunCET; \citealt{Mason2021}) instrument, which is designed to track CMEs from their initiation on the bright solar disk through their complete acceleration profile in the middle corona where they become several orders of magnitude fainter. This means that the regions of interest are nonlinear: a circle at the center of the detector for the solar disk versus the remaining area on the detector. Figure \ref{fig:shdr_median_algorithm} shows how the algorithm employs an independent but linear readout to achieve an effective nonlinear composite; in this case resulting in a short-integration-time\footnote{Note: We use the terms ``integration time" and ``exposure time" interchangeably herein.}  circular region with a long-integration-time region for the remaining area. 

 In order to remove transient noise including energetic particle hits in the data, this algorithm applies a median average across several sequential exposures. This is a well established method that is used, for example, within all of the cameras onboard the Solar Dynamics Observatory (SDO; \citealt{Pesnell2012}). This ``median image stack" and regional readout define the imaging sequence over one minute, producing a composite image that is compressed and stored with the cameras onboard processor and memory. 

Within our camera electronics software table, we can configure the duration of the long and short exposures (t$_{long}$ and t$_{short}$) and the number of each used for the median stack (N$_{long}$ and N$_{short}$). Note that this detector (and nearly all presently available similar detectors) cannot read pixels independently; it can only read rows independently. Therefore, in order to obtain the non-linear (circular) cutout at the center of the image, those rows must be read twice, with both t$_{short}$ and t$_{long}$ exposure times (green and blue, respectively, in Figure \ref{fig:shdr_median_algorithm}). This means that the blue region combines N$_{long}$-1 images for its median. As a result, N$_{long}$ should always be $\geq$4 to ensure enough images in the blue region for the median to be effective and to minimize any statistical difference between the red and blue regions. The circular region cutout and compositing occurs in camera processor software. SunCET's baseline design sets t$_{long}$=\tLong\ seconds, N$_{long}$=\nLong, t$_{short}$=\tShort\ seconds, and N$_{short}$=\nShort. These four parameters, as well as the mask that defines which pixels are in each of the three sectors (Figure \ref{fig:shdr_median_algorithm}, left), are configurable by command, allowing the flexibility to correct for any long term changes due to, e.g., solar output, instrument degradation, or systematic pointing errors. 

The orientation of the detector with respect to solar north need not stay fixed.  One composite may be oriented as in Figure \ref{fig:shdr_median_algorithm} while another is, for example, 90\degree\ rotated, meaning that the same physical area will have either 3 or 4 images in the median stack. The only difference is that the region with 4 images in the stack will have slightly less random noise. Section \ref{sec:instrument_simulator} shows signal to noise ratio contour plots, which account for this and demonstrate that the effect is negligible. But first, Section \ref{sec:lab} describes our laboratory characterization of the noise for SunCET's particular detector model and our demonstration of this algorithm. 

\section{Laboratory Validation} 
\label{sec:lab}

\begin{figure}
	\plotone{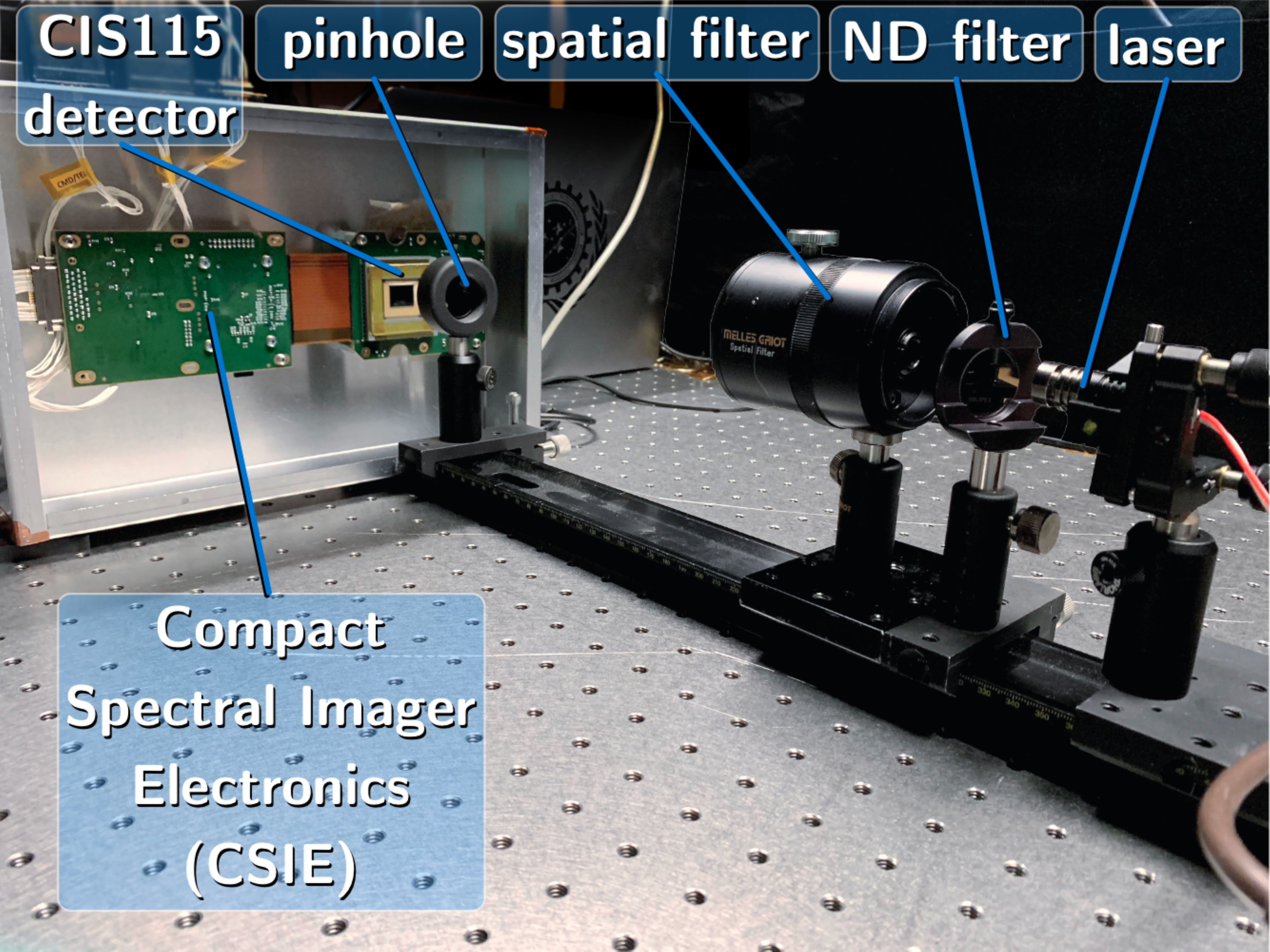}
	\caption{Laboratory optics layout to intentionally generate an Airy diffraction pattern on our detector. 
	\label{fig:lab_optics}}
\end{figure}

\begin{figure}
	\plotone{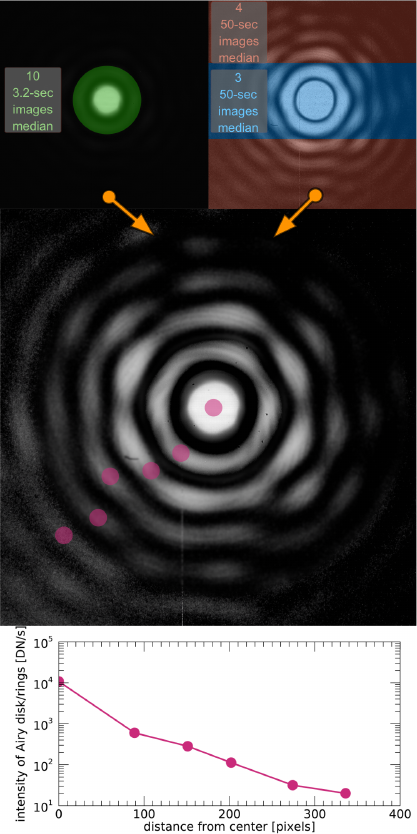}
	\caption{(top) Annotated depiction of the readout scheme applied to lab measurements of Airy diffraction patterns using SunCET's detector and electronics. (middle) Resultant composite image. The bright disk and first ring are taken at short exposure (3.2 sec) and the further rings at long exposure (50 sec). (bottom) Intensity of the disk and rings through the composite sampled at the pink circles, which shows brightness dropping dramatically, similar to the solar disk and off-disk corona. Wobbles in the rings are due to imperfections in the pinhole.
	\label{fig:airy}}
\end{figure}

In order to demonstrate that our SHDR algorithm works in practice, we ran a technology demonstration (Figure \ref{fig:lab_optics}). Light from a 405 nm laser was passed through a neutral density (ND) filter to reduce its intensity, a spatial filter to minimize distortions, and finally a 25 \textmu m diameter pinhole to produce an Airy diffraction pattern imaged onto the same model of detector (Teledyne e2v CIS115) and custom electronics intended for SunCET flight. Surrounding the bright Airy disk, each subsequent concentric Airy ring dramatically decreases in brightness (Figure \ref{fig:airy}), similar to the bright solar disk and the off-limb corona. 

We optimized the image sequence parameters for the brightness profile of this particular input source. Long exposures show the rings best, but the bright disk saturates. Short exposures prevent saturation but the signal from the rings is lost in noise. Stitching these together in the manner described in Section \ref{sec:shdr}, we obtain the composite image shown in the middle of Figure \ref{fig:airy}. The specific values of the configurable parameters were: t$_{long}$=50 sec, N$_{long}$=4, t$_{short}$=3.2 sec, and N$_{short}$=10.  

Additionally, we measured the noise characteristics of this detector, verifying the specification sheet. We measured the peak read noise (5.1  electrons), which is in good agreement with the specification (5 electrons). We also took dark exposures at room temperature and with the detector at -10 \degree C (the target flight temperature for SunCET) inside of a dark thermal chamber, finding the expected dark noise of $\sim$20 electrons/pixel/second at $\sim$20\degree C and 0.4 electrons/pixel/second at the target flight temperature.

Blooming is also a concern for the SHDR algorithm but is typically not detailed in detector specification sheets so we measured it by increasing exposure time until the laser light saturated pixels and overflowed into adjacent pixels. We found no measurable overflow beyond the immediately adjacent pixels. In cases like SunCET that bin the detector image to achieve Nyquist sampling of the telescope image resolution, this means the already minor blooming is negligible. Many other sensors of this type (CMOS) have already flown and shown no issues with blooming \citep{Seaton2013a, DeGroof2008}, for example, Project for On-Board Autonomy-2 (PROBA2; \citealt{Santandrea2013}) / Sun Watcher with Active Pixels and Image Processing (SWAP; \citealt{Seaton2013}), and Solar Orbiter (SolO; \citealt{Muller2013}) / Extreme Ultraviolet Imager (EUI; \citealt{Rochus2020}).

\begin{figure*}
	\plotone{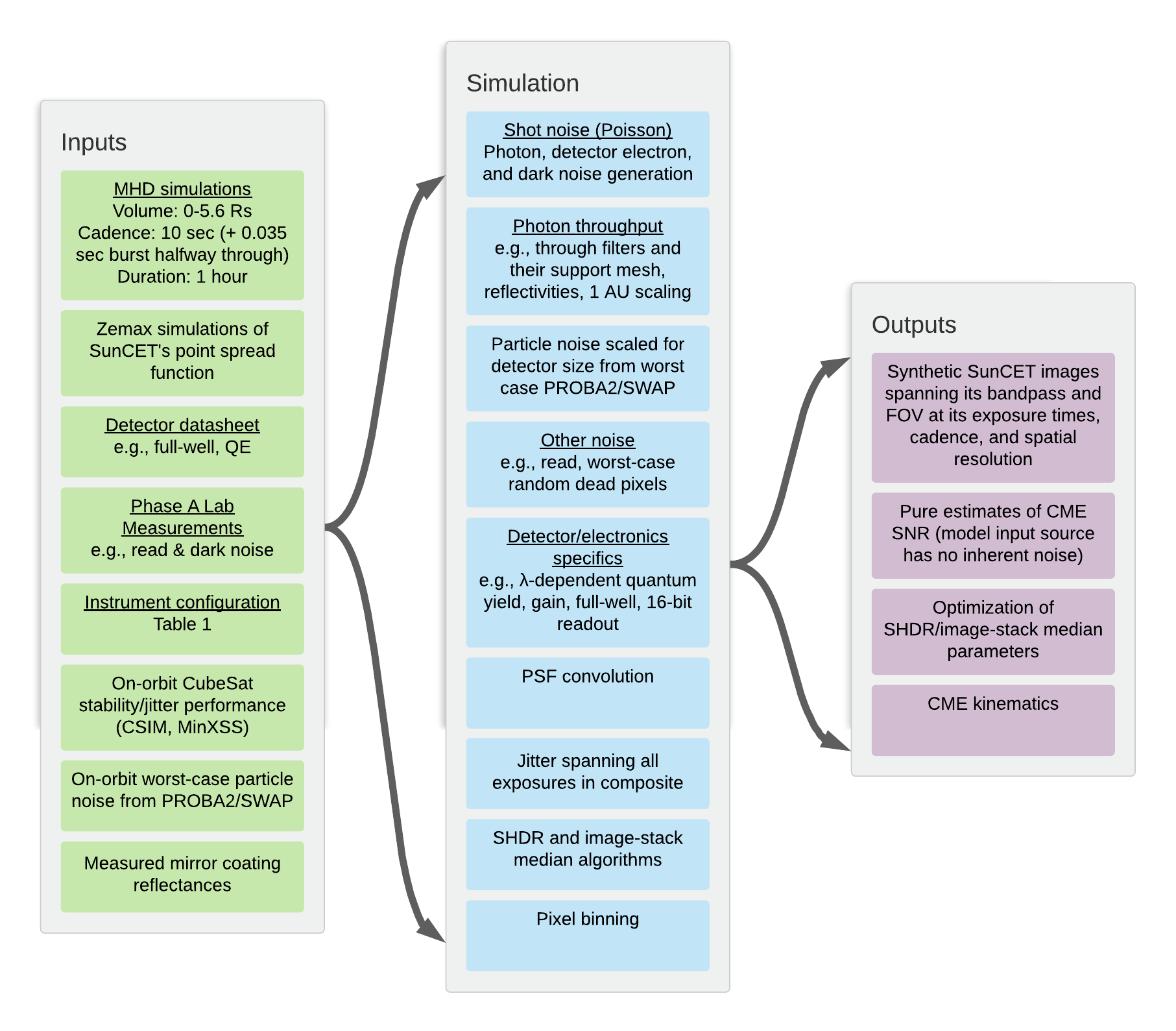}
	\caption{Summary flowchart of the SunCET instrument simulator. The order shown is not necessarily representative of the order of processing. See Table \ref{tab:instrument_response} for most quantitative values.
	\label{fig:instrument_simulator}}
\end{figure*}

Finally, timing of the components in the imaging sequence are important for synchronization, e.g., between the end of the green integrations and the end of the first red integration in Figure \ref{fig:shdr_median_algorithm}. This means that the time to clear a row in preparation for the next integration should be short. As designed and expected, we measured the row sample / reset time to be 1.5 $\mu s$, much shorter than our shortest integration time of 35000 $\mu s$. Therefore, no significant waiting time needs to be inserted to achieve synchronization throughout the imaging sequence.

\section{Instrument Simulation with Solar Extreme Ultraviolet Target} 
\label{sec:instrument_simulator}

We produced a detailed software instrument simulator to encapsulate the performance of each component of the instrument, including those verified in the lab. This tool allows us to generate synthetic images that realistically represent what flight data would look like. Figure \ref{fig:instrument_simulator} summarizes the inputs, simulation processes, and outputs; Table \ref{tab:instrument_response} quantifies crucial instrument parameters. 

\subsection{Inputs}
Inputs include magnetohydrodynamic (MHD) and instrument ray tracing simulations, component data sheet values, laboratory measurements, and relevant on-orbit performance of components.  Three MHD simulations of CMEs with a variety of speeds were run on the NASA Pleiades supercomputer following the procedure in \citet{Jin2017}, which uses Alfv\'en Wave Solar Model (AWSoM; \citealt{vanderholst2014}) within the Space Weather Modeling Framework (SWMF; \citealt{toth2012}). Section \ref{sec:lab} describes the detector verifications we made and \citet{Mason2021} present the mirror coating measurements. On-orbit pointing stability measurements for a common commercially-available attitude determination and control system (ADCS) for small satellites comes from data internal to the team for the Compact Spectral Irradiance Monitor (CSIM) and also from \citet{Mason2017} for the Miniature X-ray Solar Spectrometer (MinXSS; \citealt{MasonMinXSS2016}). We analyzed the on-orbit PROBA2/SWAP data, identifying the peak particle noise flux in its imager, which unsurprisingly occurred during a spacecraft transit through the South Atlantic Anomaly. We also included many toggles in the simulator to enable/disable, e.g., this worst-case particle noise and worst-case dead pixels.

\begin{deluxetable*}{cc}
	\tablecaption{SunCET instrument parameters.}
	\label{tab:instrument_response}
	\tablehead{\colhead{Instrument parameter} & \colhead{Value}} 
	\startdata
		Bandpass & 170 - 200 \AA \\
		Aperture size & 44.9 cm$^2$  \\ 
		Focal length & 300.42 mm  \\ 
		Magnification & 1.6x  \\ 
		Primary mirror (PM) radius of curvature & -350 mm \\ 
		PM conic constant & -1.4 \\ 
		PM outer diameter & 108 mm \\ 
		PM height$^*$ & 92.5 mm \\ 
		PM inner hole diameter & 48 mm \\ 
		Secondary mirror (SM) radius of curvature & -335 mm \\ 
		SM conic constant & -27.3 \\ 
		SM outer diameter & 48 mm \\ 
		Field of View (FOV) & 4 \Rs \\
		Entrance filter throughput & 0.6 \\
		Entrance filter mesh throughput & 0.98 \\
		Focal-plane filter throughput & 0.85 \\
		Pixel array & 1500 x 1500 \\ 
		Pixel size & 7 \textmu m x 7 \textmu m \\ 
		Plate scale & 4.8\arcsec/pixel \\ 
		$\lambda$-Averaged quantum yield & 18.3 e$^-$/ph \\ 
		Dark noise & 0.4 e$^-$/pixel/sec \\ 
		Readout noise & 5.1 e$^-$/pixel \\  
		Fano noise & 1.3 e$^-$/pixel \\ 
		Max read rate & 0.1 sec (full frame) 0.025 (up to 500 rows) \\ 
	\enddata
	\tablecomments{$^*$Circle truncated on two ends to this height}
\end{deluxetable*}

\subsection{Simulation}
The instrument simulation follows the well-established principles laid out in \textit{Photon Transfer} \citep{Janesick2007}. The boxes in the middle of Figure \ref{fig:instrument_simulator} are grouped by convenience, but the code essentially follows the photons from their source through the instrument until they are absorbed (or not) by the detector. Note that all sources of noise are handled independent of the ``pure" source signal from the MHD simulation, allowing for the clean signal to noise ratio calculations in Section \ref{sec:snr}. 

Shot noise generally follows a Poisson distribution and it occurs at several points. The generation of photons at the source in this case is primarily due to thermal excitation of electrons in ions. At the opposite end of this chain, when a photon is absorbed by the CMOS detector it generates a Poisson-distributed random number of electrons with a mean equal to the quantum yield, which is itself wavelength dependent (higher energy photons must generate more electrons on average). The quantum efficiency is the probability that the photon is absorbed at all and clearly those that aren't absorbed won't result in any electron shot noise, so they are dropped from the calculation accordingly. Finally, thermal vibrations in the detector itself release a number of electrons known as dark current or dark noise because it occurs even in the absence of incident light. This process is also random. We first generate a full size image where each pixel is assigned a Gaussian random value, with a mean and standard deviation determined from our dark noise measurements in the lab. We apply a random value from the Poisson distribution to each pixel with the mean set to whatever value was already present in the pixel.

We also include other sources of noise. We assume a larger number of dead and hot pixels than we've seen in similar detectors in flight in order to be conservative. Similarly, we use the worst case particle noise observed to date from the PROBA2/SWAP mission, which has a CMOS similar to the one being studied here. We do, however, scale the particle noise rate to account for differences in detector size, integration time, and shielding. 

There are also losses that occur as photons pass through any telescope. In this case, we have several filters and mechanical support meshes supporting them that block out visible light while letting EUV photons through, but it is not a perfect gate: some fraction of EUV photons do get absorbed. Similarly, the mirrors in our telescope have a coating that tunes the bandpass within the EUV but each photon reflection is not perfectly efficient. We also convolved the point spread function from the Zemax model with the input image. Note that for flight, the smearing effects of the point spread function will be mitigated using well-established EUV imager deconvolution methods (e.g., \citealt{Schwartz2015, Gonzalez2016, Seaton2013a}) but those enhancements are not applied in these simulations, meaning our results are conservative in this respect.

\begin{figure}
	\plotone{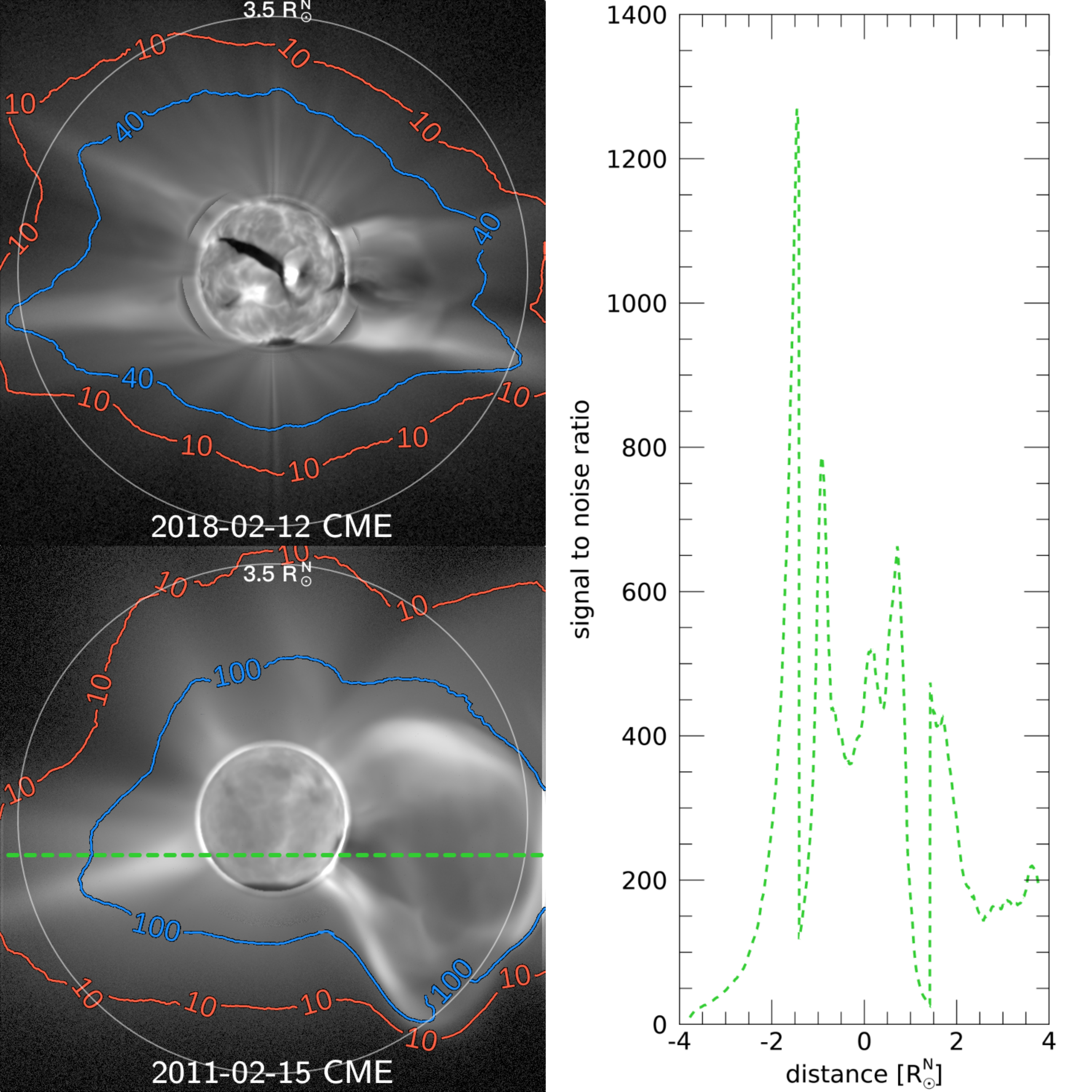}
	\caption{(left) Comparison of signal to noise ratio (SNR) contours overlaid on synthetic SunCET images for the two CME simulations that resulted in the lowest (top) and highest (bottom) SNRs. Both cases reach or exceed the International Organization for Standardization (ISO) 12232-defined ``excellent" SNR $\geq$40. (right) A trace through the 2011 CME composite image indicated by the dashed green line. 
	\label{fig:snr}}
\end{figure}

Additionally, the telescope must be mounted on a pointing platform, which will be imperfect. Our primary interest is small satellite flight so we focused on such platforms. In this case, pointing accuracy was assumed to be a nonissue given the performance of commercially available system. Stability, however, is important to simulate because it is tightly coupled with integration times, which are often a free parameter in design space to be explored and optimized. Therefore, we converted the 1$\sigma$ RMS 3U CubeSat stability adapted from \citet{Mason2017} -- verified to be consistent with results from the 6U CSIM -- to the amount of pixel shift during and across exposures.  

Finally, the detector's readout electronics and processing must be simulated. Gain is applied to amplify the signal, the full well of the sensor is accounted for as a high limit, the number of bits being read into for each row is accounted for as a high limit, pixel binning is applied, and the SHDR / image-stack median algorithms are applied. While every input and simulation component discussed here is tunable, the SHDR/median algorithm is novel and in Section \ref{sec:snr} we discuss how the variables that define its function impact the signal to noise ratio.

\subsection{Signal to Noise Ratio (SNR)}
\label{sec:snr}

By changing the exposure values of t$_{long}$, N$_{long}$, t$_{short}$, and N$_{short}$ (see Section \ref{sec:shdr}) in the simulator, we can optimize against different noise sources. Longer exposures increase signal but are more susceptible to jitter, particle spikes, and dark noise; while read and digitization noise can play a larger role in shorter exposures. Dark noise can be largely removed (up to the level of shot noise) by subtraction and particle spikes can be mostly screened by taking the median across the N exposures (Section \ref{sec:shdr}). We explored the parameter space to optimize SNR of our model CMEs, establishing the baseline values of t$_{long}$=\tLong\ seconds, N$_{long}$=\nLong, t$_{short}$=\tShort\ seconds, and N$_{short}$=\nShort. 

\begin{deluxetable*}{cccccccccc}
	\label{tab:snr}
	\tablecaption{Various configurations and resultant SNR for the 2011 CME.}
	\tablehead{\colhead{} & \colhead{N$_{long}^*$} & \colhead{t$_{long}^*$} & \colhead{N$_{short}^*$} & \colhead{t$_{short}^*$} & \colhead{Resultant} & \colhead{Pixels/Bin* /} & \colhead{SNR of CME} & \colhead{SNR of Disk}\\ 
		\colhead{} & \colhead{} & \colhead{} & \colhead{} & \colhead{} & \colhead{Cadence} & \colhead{Spatial Resolution}  & \colhead{at 3.5 \Rs} & \colhead{(mean)}} 
	\startdata
		Requirement  & -- &  $\leq$23 s & -- & $\leq$ 23 s & $\leq$192 s & $\leq$6x6 / $\leq$30\arcsec & $\geq$10 & $\geq$10\\
		Baseline &  \nLong & \tLong\ s & \nShort & \tShort\ s & 60 s & 2x2 / 9.6\arcsec  & 331 & 804\\
		Example 2 &  \nLong & 5 s & \nShort & 0.1 s & 20 s & 2x2 / 9.6\arcsec  & 186 & 1370 \\
		Example 3  & \nLong & \tLong\ s & \nShort & \tShort\ s & 60 s & 4x4 / 19.2\arcsec & 598 & 812\\
		Example 4 &  \nLong & 5 s & \nShort & 0.1 s & 20 s & 4x4 / 19.2\arcsec & 356 & 2874\\
	\enddata
	\tablecomments{* indicates parameters that are configurable in flight.}
\end{deluxetable*}

Figure \ref{fig:snr} shows the lowest and highest SNR cases from the multiple CME simulations with SNR contours overlaid on the left and on the right a horizontal trace of SNR through the middle of the 2011 CME. We performed this analysis through the entire simulation time span, from CME initiation to it reaching the edge of the FOV and found SNR $\geq$40 in all cases. Table \ref{tab:snr} shows the impact on SNR when varying different input parameters. The solution space is quite large, as evidenced by the remaining margin between these configurations and the requirements. Five of these parameters, in addition to the regional mask are even configurable in flight, meaning that there remains flexibility to meet all requirements even if on orbit performance is degraded for any reason. 

\section{Discussion} 
\label{sec:discussion}

The simultaneous high dynamic range method avoids the problems of traditional HDR for many science applications. SHDR does not require the employment of complex and expensive multiple sensors or beamsplitters and does not sacrifice temporal dynamics to do it. The number of sensors available that have independent row readout capabilities is presently relatively small but there are several commercially available today, including the one we used in our lab testing which is sensitive to both visible and ultraviolet wavelengths. Our code for the simulation presented here is also freely available for the community to use and modify for their own SHDR-parameter optimization \citep{MasonSunCETIDL2021}. The ultimate extrapolation of this work would be a per-pixel readout to enable complete control without the need for the additional linear-to-nonlinear steps required here; such detectors are starting to become commercially available.

SHDR can be applied as a solution for instruments in heliophysics, astrophysics, planetary and Earth science. We presented a case of solar spatial observation to demonstrate the technique, in particular highlighting how the linear row-based readout could be adapted into an effectively non-linear readout. A direct linear readout where each row is still controlled independently would be especially effective for spectrographs that do not introduce curvature, e.g., first order spectra: a group of rows covering each spectral line could be integrated as long as needed to achieve good SNR. Many of the spectrographic modes onboard Hubble Space Telescope produce this type of first order spectra. Future missions targeting sources with large brightness variations should include the SHDR method in their design trade studies as a solution to either increase observing efficiency or obtain temporal dynamics while maintaining low complexity in the system. 

\acknowledgments

The authors thank NASA for supporting this work as part of the SunCET Phase A study under grant 80NSSC20K1750. We also thank the entire SunCET team for helping to build the mission's science case that justified this work.

\textcolor{white}{-} % hack to prevent software from overlapping with acknowledgements
\software{
	AASTeX \citep{AASJournalsTeam2018}, 
	SunCET IDL code \citep{MasonSunCETIDL2021},
	IDL, 
	SolarSoft \citep{SolarSoft2012} ,
	IPython \citep{Perez2007}, 
	matplotlib \citep{Hunter2007}, 
	numpy \citep{Oliphant2006}, 
	pandas \citep{McKinney2010}, 
	scipy \citep{Jonesa2001},
	}

%\bibliography{/Users/jmason86/Dropbox/Research/Literature/library.bib}
\bibliography{main.bbl}{} % TODO: Update the static .bib file once no more references are being used
\bibliographystyle{aasjournal}

\listofchanges

\end{document}